\documentclass[twocolumn,showpacs,preprintnumbers,amsmath,amssymb]{revtex4}%
\usepackage{graphicx}
\usepackage{dcolumn}
\usepackage{bm}
\usepackage{color}
\usepackage{amsmath}
\usepackage{amsfonts}
\usepackage{amssymb}%
\begin{document}
\title{Wavelength Dependent Tunneling Delay Time}
\author{Xiaolei Hao$^{1}$}
\author{Zheng Shu$^{2}$}
\author{Weidong Li$^{1}$}
\email{wdli@sxu.edu.cn}
\author{Jing Chen$^{2,3}$}
\email{chen_jing@iapcm.ac.cn}
\affiliation{$^{1}$Institute of Theoretical Physics and Department of Physics, State Key
Laboratory of Quantum Optics and Quantum Optics Devices, Collaborative
Innovation Center of Extreme Optics, Shanxi University, Taiyuan 030006, China}
\affiliation{$^{2}$Institute of Applied Physics and Computational Mathematics, P. O. Box
8009, Beijing 100088, China}
\affiliation{$^{3}$HEDPS, Center for Applied Physics and Technology, Peking University,
Beijing 100084, China}
\keywords{}
\pacs{33.80.Rv, 34.50.Rk}

\begin{abstract}
A clear consensus on how long it takes a particle to tunnel
through a potential barrier has never been so urgently required,
since the electron dynamics in strong-field ionization can be
resolved on attosecond time-scale in experiment and the exact
nature of the tunneling process is the key to trigger subsequent
attosecond techniques. Here a general picture of tunneling time is
suggested by introducing a quantum travel time, which is defined as the
ratio of the travel distance to the expected value of the velocity
operator under the barrier. Specially, if applied to rectangular
barrier tunneling, it can retrieve the B\"{u}ttiker-Landauer time
$\tau_{BL}$ in the case of an opaque barrier, and has a clear
physical meaning in the case of a very thin barrier wherein
$\tau_{BL}$ can not be well defined. In the case of strong-field
tunneling process, with the help of the newly defined time, the
tunneling delay time measured by attoclock experiment can be
interpreted as a travel time spent by the electron to tunnel from a
point under the barrier to the tunnel exit. In addition, a
peculiar oscillation structure in the wavelength dependence of
tunneling delay time in deep tunneling regime is observed, which is
beyond the scope of adiabatic tunneling picture. This oscillation
structure can be attributed to the interference between the ground
state tunneling channel and the excited states tunneling channels.

\end{abstract}
\date[Date text]{\today }
\received[Received text]{\today}

\maketitle

A renewed urgency to clarify the long debating and
fundamental question in quantum mechanics
\cite{maccoll,buttiker,landauer1989,
buttiker1983,hauge1989,landauer1994,yamada,mcdonald,landsman2016,teeny,rost2016}: whether a point particle tunneling through an
energetically forbidden region takes a finite time or is instantaneous, has
been brought by the recently developed experimental capabilities in attosecond science
\cite{uiberacker,eckle1,eckle2,schultze,goulielmakis,arissian,shafir,alexandra2}%
, which can encode the tunneling time in the strong-field
ionization process and
provides significant practical implications in exploring
attosecond electron dynamics
\cite{eckle1,eckle2,schultze,goulielmakis,arissian,pfeiffer2,alexandra1}.
Especially, in the so called attoclock experiments
\cite{eckle1,eckle2,pfeiffer2,alexandra1,
barth2011,keller2012,keitel2013,gong2013,
yakaboylu2014,keitel2015,liu2016,camus} which are based on the
strong-field ionization of atoms in a laser field with
close-to-circular polarization, the photoelectron momentum
distribution at the detector is mapped into the time of the
electron appearance in the continuum, and then the time the electron
has spent to tunnel under the barrier can be extracted. Although
it seems that the attoclock experiment makes the final
solution of the long-standing tunneling time problem possible,
there have been considerable controversy in the interpretation of
the attoclock experiments\cite{alexandra1,camus,lisa,bray}. The
controversy can be classified into two categories: i) Whether the
tunneling delay time exists in principle, and how to
define and understand it. ii) How to remove the influence of the
laser field and the Coulomb field in the deconvolution procedure
to extract the exact release time from the momentum distribution.
Here we mainly focus on the problems in category i) which relate
closely to the tunneling process.

In strong-field atomic physics, the majority of phenomena
rely on the Keldysh tunneling time $\tau_{K}=\sqrt{2I_{p}}/F$
which can be obtained by calculating the classical time it takes a
particle to cross the inverted triangular barrier\cite{keldysh},
where $I_{p}$ is the electron binding potential and $F$ is the
laser electric field. If the optical period $2 \pi/\omega$ is much
longer than $\tau_{K}$, i.e., Keldysh parameter $\gamma=
\omega\tau_{K} \ll1$, the laser field can be treated as a
static electric field. In this case, the bound electron is freed
from the atom via tunneling (adiabatic process) through a potential barrier
formed by the laser field and the atomic potential. Whereas in the
case of $\gamma\gg1$, the electron sees a quickly oscillating
electric field and absorbs many photons to transit to continuum.
On the other hand, as a very popular tunneling time in the broader field of physics, B\"{u}ttiker-Landauer time
$\tau_{BL}=\int_{x_{1}}^{x_{2}}m/p(x) dx$
describes the time spent by the particle to travel from the entrance point $x_{1}$ to the exit point $x_{2}$ under the
barrier $V(x)$ with momentum $p(x)=\sqrt{2 m[V(x)-E]}$, where $E$ is the energy of the particle. $\tau_{BL}$ also characters the onset of transition from pure tunneling to
tunneling while absorbing one or more energy quanta from the
oscillating field \cite{buttiker}. So the Keldysh time and the
B\"{u}ttiker-Landauer time are very closely related not only in the
definition, but also in distinguishing the process
\cite{alexandra2}. However, these two well defined tunneling times
with clear physical meaning are ruled out as a candidate for the
tunneling delay time, since they ($\tau_{BL}\sim600$ as) are found
to deviate far from the time measured by the attoclock
experiments (less than 120 as)\cite{eckle1,bray}.

To interpret the attoclock experiments
\cite{eckle1,eckle2,pfeiffer2,alexandra1,
barth2011,keller2012,keitel2013,gong2013,
yakaboylu2014,keitel2015,liu2016,camus}, the complementary aspects
of the tunneling ionization dynamics in strong laser field have been extensively explored and
various associated times have been suggested
\cite{mcdonald,landsman2016,teeny,alexandra2,alexandra1}. The time
spent by the initial ground state in a time varying laser field to
develop the under barrier wave function components necessary for
reaching a static field ionization rate was suggested as tunneling
time in Ref. \cite{mcdonald}. Based on the Feynman path integral
approach, authors in Refs. \cite{alexandra2,alexandra1} calculated
the probability distribution of tunneling times and found that the
common tunneling time definitions can be viewed as averaged quantities
rather than deterministic values. By numerically solving the
time-dependent Schr\"{o}dinger equation (TDSE) and employing a
virtual detector at the tunnel exit, a finite positive time delay
between the electric field maximum and the instant of ionization
was identified in Ref. \cite{teeny}. And the following attoclock
measurement on two atomic species with slightly deviating atomic
potentials \cite{camus} supported this nonzero tunneling delay
time.

Considering that momentum operator is well
defined in quantum mechanics, here we introduce a naive
definition of quantum travel time by analogy with the classical travel time
\begin{equation}
\tau_{t}=\frac{m\left\vert \mathbf{x}_{2}-\mathbf{x}_{1}\right\vert }{\left\vert
|\bar{\mathbf{p}}|\right\vert }, \label{tabs}%
\end{equation}
where $m$ is the mass of the particle, $\left\vert
|\cdots|\right\vert $ means making the modula of \textquotedblleft$\cdots$", and
$\bar{\mathbf{p}}$ is the average momentum of the particle during its staying within the region
$({x}_{1}, {x}_{2})$ and can be obtained by calculating the expected value of the momentum operator
$\hat{\mathbf{p}}=-i\hbar\mathbf{\triangledown}$
\begin{equation}
\bar{\mathbf{p}}=\frac{\int_{\mathbf{{x}_{1}}}^{\mathbf{{x}_{2}}}\psi^{\ast}(\mathbf
{x})\hat{\mathbf{p}}\psi(\mathbf{x})d\mathbf{x}}{\int_{\mathbf{{x}_{1}}}^{\mathbf{{x}_{2}}}\psi^{\ast}(\mathbf{x})\psi(\mathbf{x})d\mathbf{x}} \label{v}%
\end{equation}
with $\psi(\mathbf{x})$ the wave function of the
particle within the region $({x}_{1}, {x}_{2})$. The quantum travel time in Eq. (\ref{tabs}) tells us how long it takes a particle to travel through two
spatially separated positions $\mathbf{x}_{1}$ and
$\mathbf{x}_{2}$ in quantum mechanics, and can be
considered as tunneling time if $\psi(\mathbf{x})$ is the wave function under the barrier. Note that $\tau_{t}$ is also an average
quantity rather than a deterministic one. In this letter, first we
check the validity and rationality of the definition about
$\tau_{t}$ in the rectangular barrier tunneling problem. Then we
apply it to the strong-field tunneling process to interpret the
tunneling delay time measured by attoclock experiment. Finally, as
an example of applying attoclock technique to explore attosecond
electron dynamics, we identify a coherent tunneling process by
studying the wavelength dependence of tunneling delay time.

\begin{figure}[ptb]
\centering \vspace{-0 in}\includegraphics[width=3.3 in]{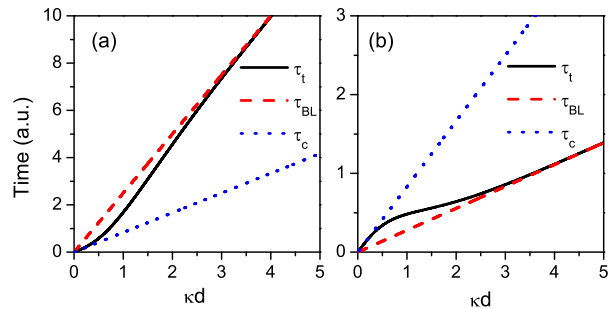} \vspace{-0.1
in}\caption{(Color online) The quantum travel time $\tau_{t}$, the
B\"{u}ttiker-Landauer time $\tau_{BL}$ and the free travel time
$\tau_{c}$ as functions of $\kappa d$ with (a) $E=1.8$ a.u., $V_{0}=2.0$ a.u. and (b)
$E=0.2$ a.u., $V_{0}=2.0$ a.u.. The atomic units are used with
$e=m=\hbar=1$.}%
\label{fig1}%
\end{figure}

\textit{Rectangular barrier tunneling} Considering a particle with kinetic energy $E=\hbar^{2}k^{2}/2m$
moving along the $x$ axis and interacting with a rectangular barrier of height
$V_{0}$ and width $d$ centered at $x=0$, the average momentum of the particle within the barrier is (see textbook on quantum mechanics, for example,
Ref. \cite{landau})
\begin{equation}
\overline{p}=R\left(  r,\triangle\theta\right)  \hbar k+iI\left(
r,\triangle\theta\right)  \hbar\kappa \label{vp}%
\end{equation}
with%
\begin{equation}
R\left(  r,\triangle\theta\right)=\frac{2\kappa^{2}}{\kappa^{2}-k^{2}%
}\frac{\left(  \kappa^{2}+k^{2}\right)  rd}{\left(  \kappa^{2}+k^{2}\right)
rd-2k\cot(\triangle\theta)}\label{re}
\end{equation}
and
\begin{equation}
I\left(  r,\triangle\theta\right)=\frac{2k\kappa}{\kappa^{2}-k^{2}}%
\frac{2k}{\left(  \kappa^{2}+k^{2}\right)  rd-2k\cot(\triangle\theta)},
\label{im}%
\end{equation}
where $\kappa=\sqrt{2m\left[  V_{0}-E\right]  }/\hbar$, $r$ is the ratio of
transmission probability to reflection probability, and $\triangle\theta$
is the phase increased across the barrier. In the limit of the very thin
barrier ($\kappa d\ll1$), we have $R\left(  r,\triangle\theta\right)  \approx1$ and $I\left(
r,\triangle\theta\right)  \approx\kappa^{-2}\left(  \kappa^{2}+k^{2}\right)
\kappa d\ll1$, which means that the movement of the tunneling particle can be
treated as a free one with $\bar{p}=\hbar k$.
While in the opposite limit of opaque barrier ($\kappa d\gg1$), we simply have $R\left(  r,\triangle\theta\right)  \approx0$ and
$I\left(  r,\triangle\theta\right)  \approx1$. The momentum of the
tunneling particle is imaginary and approaches the effective
Buttiker-Landauer momentum $\hbar\kappa$ \cite{buttiker}.

The quantum travel (or tunneling) time $\tau_{t}=md/\hbar\sqrt{R^2 k^2+I^2\kappa^2}$ is inversely proportional to the average momentum, and its dependence on $\kappa d$ is shown in Fig.~\ref{fig1}. The Buttiker-Landauer time $\tau_{BL}$
and the free travel time $\tau_{c}$ which denotes the time a free particle spends traveling through the same distance but without
potential barrier, are also shown for comparison. Interesting thing
is that $\tau_{t}$ is always bounded by two limit cases, free
travel time $\tau_{c}$ for very thin barrier and
B\"{u}ttiker-Landauer time $\tau_{BL}$ for opaque barrier.
Therefore, $\tau_{t}$ not only can retrieve the
B\"{u}ttiker-Landauer time $\tau_{BL}$ in the case of opaque
barrier, but also has a clear meaning even in the case of very
thin barrier, wherein $\tau_{BL}$ cannot be well defined
\cite{hauge1989,landauer1994}.

\begin{figure}[ptb]
\centering \vspace{-0 in}\includegraphics[width=3.3 in]{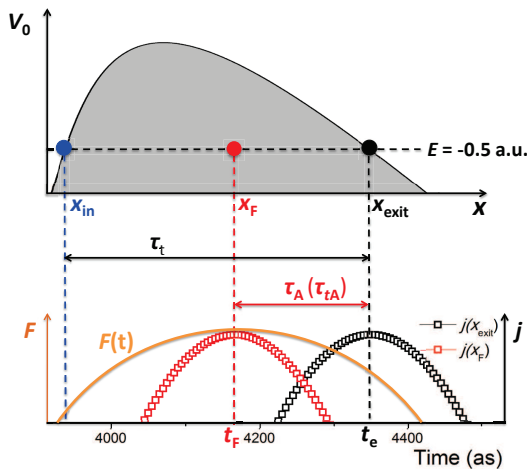} \vspace{-0.1
in}\caption{(Color online) The diagram of relation between the tunneling delay time
$\tau_{A}$, the quantum travel times $\tau_{t}$ and $\tau_{tA}$. Extraction of $\tau_{A}$ is realized by comparing the peak
of the probability current $j(x_{exit},t)$ at $t_{e}$ with the
peak of the laser field $F(t)$ at $t_{F}$. $x_{F}$ is the position at which the peak of $j(x_{F},t)$
coincides with the peak of $F(t)$. $\tau_{t}$
describes the time spent by electron to travel from $x_{in}$ to $x_{exit}$, while $\tau_{tA}$ corresponds to the travel region from $x_{F}$ to $x_{exit}$. The laser parameters is $F_{0}=0.04$ a.u. and
the wavelength is $5 \mu m$.}%
\label{fig2}%
\end{figure}

\textit{Tunneling delay time in attoclock measurement} The
interpretations of the attoclock experiments are mainly based on
the following assumptions \cite{lisa}: i) the highest probability
for the electron to tunnel is at the peak of the electric field;
ii) ionization is completed once the electron emerges from the
barrier; iii) after the barrier exit, the electron dynamics is
classical. The first two assumptions are closely related to the
tunneling process, and imply that the experimentally measured
delay time $\tau_{A}$, in principle, is the time interval between
the instant of the peak of the laser field (as a reference instant
$t_{F}$) and the instant at which the ionized electron wave packet
appears at the tunnel exit. Theoretically, the instant the electron leaves the barrier can be read from the probability current, so $\tau_{A}$ can
be directly extracted from our calculation. By comparing it with
the quantum travel time $\tau_{t}$ defined in Eq. (\ref{tabs}) as
well as the B\"{u}ttiker-Landauer time $\tau_{BL}$, the physical
meaning of $\tau_{A}$ can be clarified.

Tunneling of an initially bounded electron through the $3$D
Coulomb potential bent by a linearly polarized intense laser
field, can be reduced to an effective 1D tunneling
process via introducing parabolic coordinates \cite{landau}.
Therefore, the tunneling process in linearly polarized intense laser
field can be well
described by $1$D TDSE (atomic units are used below with
$e=m=\hbar=1$)
\begin{equation}
i\frac{\partial\psi(x,t)}{\partial t}=\left[  -\frac{1}{2}\frac{\partial^{2}%
}{\partial x^{2}}-\frac{1}{\sqrt{x^{2}+\alpha}}-F(t)x\right]  \psi(x,t),
\label{schrodinger}%
\end{equation}
where $F(t)=F_{0}\sin(\omega t)$ is the laser electric field with frequency
$\omega$, $\alpha$ is a soften parameter which is chosen to be
$2.0$ for a hydrogen atom.

In our calculation, the instant at which the tunneling
electron arrives at a fixed position $x_{0}$ is the instant at
which the probability current $j(x_{0},t)$ reaches its maximum. As
shown in Fig. \ref{fig2}, the escape instant can be extracted by
reading the maximum of $j(x_{exit},t)$, and the tunneling delay
time $\tau_{A}$ is the time interval between the escape instant $t_{e}$
and the reference instant $t_{F}$ \cite{teeny}. On the
other hand, we can also determine the location of tunneling
electron at $t_{F}$ by scanning $j(x,t)$ at different positions.
As shown in Fig. \ref{fig2}, we can find a position $x_{F}$ under
the barrier, at which the peak of $j(x_{F},t)$ coincides with the
peak of the laser field $F(t)$.

\begin{figure}[pt]
\centering \vspace{-0 in}\includegraphics[width=3.3 in]{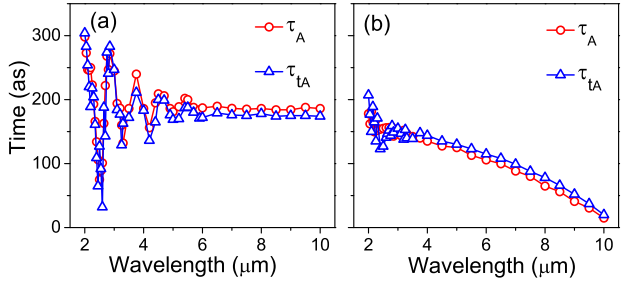}
\vspace{-0.1 in}\caption{(Color online) Wavelength dependence of
tunneling delay time $\tau_{A}$ and quantum travel time $\tau_{tA}$ for different
laser fields: (a) $F_{0}= 0.04$ a.u. and (b)
$F_{0}= 0.05$ a.u..}%
\label{fig3}%
\end{figure}

Combining with the knowledge of travel region and wave function
under the barrier, the quantum travel time can be obtained by Eq.
(\ref{tabs}). In Fig. \ref{fig2}, there are two quantum travel times
$\tau_{t}$ and $\tau_{tA}$ corresponding to the travel regions
$(x_{in}, x_{exit})$ and $(x_{F}, x_{exit})$, respectively.
Apparently,  it is $\tau_{tA}$, which is much shorter than
$\tau_{t}$, that relates closely to the tunneling delay time
$\tau_{A}$. Since both of the B\"{u}ttiker-Landauer time
$\tau_{BL}$ and the Keldysh time $\tau_{K}$ are defined in the
same travel region as $\tau_{t}$, they are found to deviate far
from $\tau_{A}$. To further illuminate the properties of these times,
we have performed 1D TDSE calculations for two different
intensities of laser field with a relative wide wavelength range,
which lie in the deep tunneling regime with Keldysh parameter
$\gamma< 0.6$. Three points deserve to be emphasized. i) The travel time $\tau_{t}$ depends only on the intensity of laser field,
but not on the wavelength, the same as the
B\"{u}ttiker-Landauer time $\tau_{BL}$ does. ii)
$\tau_{t}$ is shorter than $\tau_{BL}$. For example,
$\tau_{BL}=650$ as $\&$ $\tau_{t}=400$ as for $F_{0}=0.04$ a.u.,
and $\tau_{BL}=560$ as $\&$ $\tau_{t}=305$ as for $F_{0}=0.05$
a.u.. iii)
There is a well consistence between the quantum travel time
$\tau_{tA}$ and the tunneling delay time $\tau_{A}$, both of which
are much shorter than $\tau_{t}$ and $\tau_{BL}$, in a wide
wavelength range as can be seen in Fig. \ref{fig3}.
Therefore, the tunneling delay time measured by attoclock
experiment can be interpreted as the time spent by the electron to
tunnel from a point under barrier ($x_{F}$) to the tunnel
exit, which would be much shorter than the B\"{u}ttiker-Landauer time
$\tau_{BL}$ and the Keldysh time $\tau_{K}$.

\textit{Wavelength dependent tunneling delay time}  According to the
adiabatic tunneling theory, the tunneling delay time will be
determined only by the intensity of the laser field. This has
partially inspired a series of experiments to measure the
dependence of the tunneling delay time on laser intensity
\cite{alexandra1}. However, we find that the wavelength dependence
of tunneling delay time $\tau_{A}$ exhibits a peculiar oscillation
even in the deep tunneling regime ($\gamma<0.6$), which is beyond
the scope of the adiabatic tunneling picture. As shown in Fig.
\ref{fig3}, when the amplitude of the laser field is fixed to
$F_{0}=0.04$ a.u., $\tau_{A}$ shows a quick oscillation with
decreasing amplitude, and finally approaches to a constant. While
increasing the laser field to $F_{0}=0.05$ a.u., the oscillation
in short wavelength region becomes weak and $\tau_{A}$ decreases
monotonously with wavelength. Even a negative tunneling delay time
$\tau_{A}$ can be found when wavelength is over 10 $\mu$m (not
shown here). The decrease of $\tau_{A}$ can be attributed to the depletion of the ground state: a
loss of population before the peak of the field would enhance the
relative contribution of early ionization events \cite{lisa}. The
travel time $\tau_{tA}$ is also shown in Fig. \ref{fig3} for
comparison and is found in coincidence with $\tau_{A}$ for all
laser parameters considered here.

\begin{figure}[ptb]
\centering \vspace{-0 in}\includegraphics[width=3.3 in]{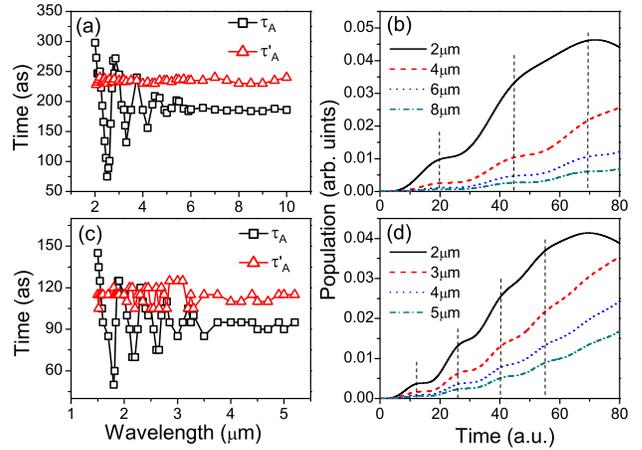} \vspace{-0.1
in}\caption{(Color online) (a) and (c): wavelength dependence of tunneling delay times $\tau_{A}$ and $\tau^{\prime}_{A}$.  $\tau^{\prime}_{A}$ is calculated by removing the component of the first excited state in the total wave function at every step of temporal evolution. (b) and (d): temporal evolution of the population of the first excited state at different wavelengths. (a) and (b) are calculated for Coulomb potential at $F_{0}= 0.04$ a.u., while (c) and (d) are calculated for quantum dot potential with only two bounded states involved at $F_{0}= 0.08$ a.u.. }%
\label{fig4}%
\end{figure}

On close inspection of Fig. \ref{fig3}(a), it is surprised that the
oscillation keeps a constant period around 850 nm. We will see below
that the oscillation is actually a result of interference between the
ground state tunneling channel and the excited states tunneling
channels. First we perform a calculation in which the component of
the first excited state in the total wave function is removed at every step
of temporal evolution. After doing so, as shown in Fig. \ref{fig4}(a), the oscillation disappears and an almost constant tunneling delay time $\tau^{\prime}_{A}$ is found, which is larger than $\tau_{A}$
at long wavelength. This is a clear evidence that the first
excited state plays an important role in the construction of the
oscillation. Then we investigate the temporal evolution of the
population of the first excited state at different
wavelengths. As shown in Fig. \ref{fig4}(b), the population shows a
series of steps with equal time intervals close to
$2\pi/\Delta E$, where $\Delta E=0.2671$ a.u. is the energy gap
between the ground state and the first excited state in our model. And the position of each step in Fig. \ref{fig4}(b) is independent of wavelength.
This kind of dynamical feature can be well understood by considering a
two-level system with dynamical coupling when the photon energy of
laser is much smaller than the energy gap (see supplementary for
details). Since the first step is usually the most prominent, at time
$t_{s}$ when the first step emerges, a considerable amount of wave
function is excited to the first excited state and then evolves
until tunneling at $T/4$ ($T$ is the optical cycle). Tunneling current through the above
channel will interfere with that direct from the ground state. The phase
difference between these two channels can be simply estimated by $\Delta
\varphi=\Delta E\left( T/4-t_{s}\right) $. Since $\Delta E$ and $t_{s}$ are independent of the optical cycle $T$, a phase shift $\Delta \varphi$ of $2\pi$ corresponds to change of the optical cycle as $8\pi/\Delta E$, which is nothing else but the period of the interference shown in Fig. \ref{fig3}(a).  After converting optical cycle to wavelength, the estimated value of the oscillation period
is determined as 682 nm which is smaller than that read from Fig.
\ref{fig3}(a). This difference is not surprising since the
contributions of the higher excited states are not included in the
above analysis.

In order to exclude the effects of the higher excited states, we
employ a quantum dot potential with only two bounded states
with energy gap $\Delta E=0.4328$ a.u. (see supplementary material for more details). As shown
in Fig. \ref{fig4}(c), similar to the case of Coulomb potential,
obvious oscillation exists in the wavelength dependence of
$\tau_{A}$ and almost disappears if the component of the
first excited state in the wave function is removed at every step
of temporal evolution. The population of the first excited state
also shows a series of steps with constant time interval
determined by the energy gap. The period of the
oscillation is estimated as 421 nm which is very close to the
period (430 nm) read from Fig. \ref{fig4}(c). Successful prediction
of the oscillation period provides a strong support to the
interpretation of the oscillation as a result of the interference
between the ground state and the excited states tunneling channels.

It is noteworthy that, when the wavelength or the intensity of the laser
field increases, the step structure in the population of the first
excited state becomes weaker (see Figs. \ref{fig4}(b),(d) and the
supplementary materials), which results in a decreasing amplitude
of the oscillation with increasing wavelength (Figs. \ref{fig3}(a)) and intensity of the laser field (Fig.
\ref{fig3}(b)).

In conclusion, based on a newly introduced quantum travel time,
the tunneling delay time measured by attoclock experiment can be interpreted
as the travel time spent by the electron to tunnel from a point under
barrier to the tunnel exit, which is actually a part of the B\"{u}ttiker-Landauer time. Our interpretation may bridge the gap between the conventional tunneling time (B\"{u}ttiker-Landauer time and Keldysh time) and the measured tunneling delay time. In addition, a peculiar oscillation structure with constant period in
the wavelength dependence of tunneling delay time is observed in
deep tunneling regime, which is beyond the scope of the adiabatic
tunneling picture. This oscillation structure can be
attributed to the interference between the ground state tunneling
channel and the excited states tunneling channels. Our results reveal
the important role of the excited states in strong-field tunneling
process, which is usually ignored. On the other hand, the identified coherent
tunneling process paves the way towards probing and
imaging of the tunneling dynamics of excited states.

X. Hao and Z. Shu contributed equally to this work. This work was
supported by the National Key Research and
Development program (No. 2016YFA0401100) and NNSFC (Nos. 11334009,
11425414, 11504215, and 11874246).


\begin{thebibliography}{99}                                                                                               %
\bibitem {maccoll}L. A. MacColl, Phys. Rev. \textbf{40}, 621 (1932).

\bibitem {buttiker}M. B\"{u}ttiker and R. Landauer, Phys. Rev. Lett.
\textbf{49}, 1739 (1982).

\bibitem {landauer1989}R. Landauer, Nature (London) \textbf{341}, 567 (1989).

\bibitem {buttiker1983}M. B\"{u}ttiker, Phys. Rev. B \textbf{27}, 6178 (1983).

\bibitem {hauge1989}E. H. Hauge, J.A. St{\o}vneng, Rev. Modern Phys. \textbf{61}, 917, (1989).

\bibitem {landauer1994}R. Landauer, T. Martin, Rev. Modern Phys. \textbf{66}
(1994) 217.

\bibitem {yamada}N. Yamada, Phys.
Rev. Lett. \textbf{93}, 170401 (2004).

\bibitem {mcdonald}C. R. McDonald, G. Orlando, G. Vampa, and T. Brabec, Phys.
Rev. Lett. \textbf{111}, 090405 (2013).

\bibitem {landsman2016}T. Zimmermann, S. Mishra, B. R. Doran, D. F. Gordon, and A. S. Landsman, Phys.
Rev. Lett. \textbf{116}, 233603 (2016).

\bibitem {teeny}N. Teeny, E. Yakaboylu, H. Bauke, and C.
H. Keitel, Phys. Rev. Lett. \textbf{116}, 063003 (2016).

\bibitem {rost2016}H. Ni, U. Saalmann, and J.-M. Rost, Phys. Rev. Lett. \textbf{117}, 023002 (2016).

\bibitem {uiberacker}M. Uiberacker, T. Uphues, M. Schultze, A. J. Verhoef, V. Yakovlev, M. F. Kling, J. Rauschenberger, N. M. Kabachnik, H. Schr\"{o}der, M. Lezius, K. L. Kompa, H.-G. Muller, M. J. J. Vrakking, S. Hendel, U. Kleineberg, U. Heinzmann, M. Drescher and F. Krausz, Nature (London) \textbf{446}, 627 (2007).

\bibitem {eckle1}P. Eckle, A. N. Pfeiffer, C. Cirelli, A. Staudte, R.
D\"{o}rner, H. G. Muller, M. B\"{u}ttiker, and U. Keller, Science \textbf{322}, 1525 (2008).

\bibitem {eckle2}P. Eckle, M. Smolarski, P. Schlup, J. Biegert, A. Staudte, M.
Sch\"{o}ffler, H. G. Muller, R. D\"{o}rner, and U. Keller, Nat. Phys. \textbf{4}, 565 (2008).

\bibitem {schultze}M. Schultze, M. Fie{\ss}, N. Karpowicz, J. Gagnon, M. Korbman, M. Hofstetter, S. Neppl, A. L. Cavalieri, Y. Komninos, T. Mercouris, C. A. Nicolaides, R. Pazourek, S. Nagele, J. Feist, J. Burgd\"{o}rfer, A. M. Azzeer, R. Ernstorfer, R. Kienberger, U. Kleineberg, E. Goulielmakis, F. Krausz and V. S. Yakovlev, Science \textbf{328},
1658 (2010).

\bibitem {goulielmakis}E. Goulielmakis, Z.-H. Loh, A. Wirth, R. Santra, N. Rohringer, V. S. Yakovlev, S. Zherebtsov, T. Pfeifer, A. M. Azzeer, M. F. Kling, S. R. Leone and F. Krausz, Nature \textbf{466}, 739 (2010).

\bibitem {arissian}L. Arissian, C. Smeenk, F. Turner, C. Trallero, A.V.
Sokolov, D. M. Villeneuve, A. Staudte, and P. B. Corkum, Phys. Rev. Lett. \textbf{105},
133002 (2010).

\bibitem {shafir}D. Shafir, H. Soifer, B. D. Bruner, M. Dagan, Y. Mairesse, S. Patchkovskii, M. Y. Ivanov, O. Smirnova and N. Dudovich, Nature \textbf{485}, 343 (2012).

\bibitem {alexandra2}A. S. Landsman and U. Keller, Phys. Rep. \textbf{547}, 1 (2015).

\bibitem {pfeiffer2}A. N. Pfeiffer, C. Cirelli, M. Smolarski, D. Dimitrovski, M. A.-Samha, L. B. Madsen and U. Keller, Nature
Phys. \textbf{8}, 76 (2012).

\bibitem {alexandra1}A. S. Landsman, M. Weger, J. Maurer, R. Boge, A. Ludwig,
S. Heuser, C. Cirelli, L. Gallmann, and U. Keller, Optica \textbf{1}, 343 (2014).

\bibitem {barth2011}I. Barth and O. Smirnova, Phys. Rev. A \textbf{84}, 063415 (2011).

\bibitem {keller2012}A. N. Pfeiffer, C. Cirelli, A. S. Landsman, M. Smolarski, D. Dimitrovski, L. B. Madsen, and U. Keller, Phys. Rev. Lett. \textbf{109}, 083002 (2012).

\bibitem {keitel2013} M. Klaiber, E. Yakaboylu, H. Bauke, K. Z. Hatsagortsyan, and C. H. Keitel, Phys. Rev. Lett. \textbf{110}, 153004 (2013).

\bibitem {gong2013} M. Li, Y. Liu, H. Liu, Q. Ning, L. Fu, J. Liu, Y. Deng, C. Wu, L.-Y. Peng, and Q. Gong, Phys. Rev. Lett. \textbf{111}, 023006 (2013).

\bibitem {yakaboylu2014} E. Yakaboylu, M. Klaiber, and K. Z. Hatsagortsyan, Phys. Rev. A \textbf{90}, 012116 (2014).

\bibitem {keitel2015} M. Klaiber, K. Z. Hatsagortsyan, and C. H. Keitel, Phys. Rev. Lett. \textbf{114}, 083001 (2015).

\bibitem {liu2016} M. Li, J.-W. Geng, M. Han, M.-M. Liu, L.-Y. Peng, Q. Gong, and Y. Liu, Phys. Rev. A \textbf{93}, 013402 (2016).


\bibitem {camus}N. Camus, E. Yakaboylu, L. Fechner, M.
Klaiber, M. Laux, Y. H. Mi, K. Z. Hatsagortsyan, T. Pfeifer,
C. H. Keitel, and R. Moshammer, Phys. Rev. Lett \textbf{119},
023201 (2017).

\bibitem {lisa}L. Torlina, F. Morales, J. Kaushal, I. Ivanov, A. Kheifets, A. Zielinski, A. scrinzi, H. G. Muller, S. Sukiasyan, M. Ivanov and O. Smirnova, Nature Phys. \textbf{11}, 503 (2015).

\bibitem {bray}A. W. Bray, S. Eckart and A. S. Kheifets, Phys. Rev. Lett \textbf{121},
123201 (2018).

\bibitem {keldysh}L. V. Keldysh, Sov. Phys. JETP \textbf{20}, 1307 (1965).

\bibitem {landau}L. D. Landau and E. M. Lifshitz, \emph{Quantum Mechanics}
(Pergamon Press, New York, 1977).


\end{thebibliography}
\end{document}